\documentstyle[12pt,fleqn]{article}
\input psfig
\def\be{\begin{equation}}
\def\ee{\end{equation}}
\def\bea{\begin{eqnarray}}
\def\eea{\end{eqnarray}}
\def\la{\mathrel{\mathpalette\fun <}}
\def\ga{\mathrel{\mathpalette\fun >}}
\def\fun#1#2{\lower3.6pt\vbox{\baselineskip0pt\lineskip.9pt
        \ialign{$\mathsurround=0pt#1\hfill##\hfil$\crcr#2\crcr\sim\crcr}}}
\def\re#1{{[\ref{#1}]}}
\def\ret#1#2{{[\ref{#1},\ref{#2}]}}

\def\thesection{}

\begin{document}
\thispagestyle{empty}
\input psfig
\def\pag{{{\langle P \rangle}}}
\null\vspace{-42pt}
\begin{flushright}
\baselineskip=12pt
{\footnotesize
FERMILAB--PUB--93/335--A\\
astro-ph/9311037\\
November, 1993}
\end{flushright}
\renewcommand{\thefootnote}{\fnsymbol{footnote}}
\baselineskip=24pt

\vspace{42pt}
\begin{center}
{\Large \bf Non-linear axion dynamics and formation
\\          of cosmological pseudo-solitons}\\
\vspace{1.0cm}
\baselineskip=14pt

Edward W.\ Kolb\footnote{Electronic mail: rocky@fnas01.fnal.gov}\\
{\em NASA/Fermilab Astrophysics Center, \\
Fermi National Accelerator Laboratory, Batavia, IL~~60510, and\\
Department of Astronomy and Astrophysics, Enrico Fermi Institute\\
The University of Chicago, Chicago, IL~~ 60637}\\
\vspace{0.4cm}
Igor I.\ Tkachev\footnote{Electronic mail: tkachev@fnas13.fnal.gov}\\
{\em NASA/Fermilab Astrophysics Center\\
Fermi National Accelerator Laboratory, Batavia, IL~~60510, and\\
Institute for Nuclear Research of the Academy of Sciences of
Russia\\Moscow 117312, Russia}\\
\end{center}

\baselineskip=24pt

\begin{quote}
\hspace*{2em} The $(3+1)$-dimensional evolution of an inhomogeneous axion field
configuration around the QCD epoch is studied numerically, including
important non-linear effects due to the attractive self-interaction.
It is found that axion perturbations on scales corresponding to causally
disconnected regions at $T \sim 1 \, {\rm GeV}$ can lead to very dense
pseudo-soliton configurations we call axitons.  These configurations evolve to
axion miniclusters with present density $\rho_a \ga
10^{-8}\,{\rm g \, cm^{-3}}$. This is high enough for the collisional
$2a \rightarrow 2a$ process to lead to Bose--Einstein relaxation in
the gravitationally bound clumps of axions, forming Bose stars.
\vspace*{12pt}

PACS number(s): 98.80.Cq, 14.80.Gt, 05.30.Jp, 98.70.-f

\end{quote}
\newpage
\baselineskip=24pt
\setcounter{page}{1}
\renewcommand{\thefootnote}{\arabic{footnote}}
\addtocounter{footnote}{-2}

\thesection{\centerline{\bf I. INTRODUCTION}}
\setcounter{section}{1}
\setcounter{equation}{0}
\vspace{18pt}

The invisible axion is one of the best motivated candidates for cosmic
dark matter. The axion is the pseudo-Nambu--Goldstone boson resulting
from the spontaneous breaking of a $U(1)$ global symmetry known as
the Peccei--Quinn, or PQ, symmetry.  The PQ symmetry is introduced to
explain the apparent smallness of strong CP-violation in QCD \re{pq}.
Although there are other possible solutions to the strong CP problem
\re{others}, and the origin of the axion in the breaking of a {\em global}
symmetry has been criticized \re{global}, the axion remains the best
known cure for the disease of strong-CP violation.

There are stringent astrophysical \ret{ac1}{ac2} and cosmological
\re{cc} constraints on the properties of the axion. In particular, the
combination of cosmological and astrophysical considerations restrict
the axion decay constant $f_a$ and the axion mass $m_a$ to be in the
narrow windows $10^{10}\, {\rm GeV} \la f_a \la 10^{12}\, {\rm GeV}$,
and $10^{-5}\, {\rm eV} \la m_a \la 10^{-3}\, {\rm eV}$ \re{mt90}.
The contribution to the mean density of the Universe from axions with
mass in this window is guaranteed to be cosmologically significant.
Thus, if axions exist, they will be dynamically important in the
present evolution of the Universe.

In addition to the usual role in the evolution of primordial density
fluctuations and the formation of large-scale structure common to all
cold dark matter candidates, axions have unique features as dark
matter.  The energy density in axions corresponds to coherent scalar
field oscillations, driven by a displacement of the initial value of
the field (the ``misalignment'' angle) away from the eventual minimum
of the temperature-dependent potential. During the QCD epoch
fluctuations in the misalignment angle on scales comparable to the
Hubble radius at that time \re{mt86} are transformed into large
amplitude density fluctuations, which later lead to tiny
gravitationally bound ``miniclusters'' \re{hr88}. It was found that
the density in miniclusters exceeds by ten orders of magnitude the
local dark matter density in the Solar neighborhood \re{hr88}. This
might have a number of astrophysical consequences, as well as
implications for laboratory axion searches \re{ps83}.

In previous studies of the evolution of the axion field around the QCD
epoch, the effect of spatial gradients of the axion field were either
neglected, or were included in a limit
where the non-linear  potential was approximated by a {\em linear} harmonic
potential.  Both approximations are
adequate for temperatures well below the QCD scale where the coherent
axion oscillations can be treated as pressureless, cold dust.
However, in a previous paper \re{kt93} we found that just at the
crucial time when the inverse mass of the axion is approximately the
size of the Hubble radius and fluctuations of misalignment angle are
still of order $\pi$, both the non-linear interaction and the gradient
terms are important, and a full field-theoretical calculation is
needed.  Here, we present the results of a 3-dimensional numerical
study of the evolution of the inhomogeneous axion field around the QCD
epoch. We find that the resulting axion clumps are much denser than
previously thought, even reaching the critical conditions for Bose
star formation \re{it91}.

In Sec.\ II we review the basic scenario for the evolution of the axion
field around the QCD epoch.  In Sec.\ III, after deriving the
equations of motions in a suitable form, we present the results of
($3+1$)-dimensional numerical calculations of an initial white-noise
axion distribution.  We find that the non-linear potential leads to the
formation of dense, roughly spherical, soliton-like axion
configurations we call axitons.  We then follow the subsequent
evolution of these spherically symmetric configurations in a
($1+1$)-dimensional calculation.  Sec.\ IV is devoted to the
consideration of an initial axion field that results in a network of
topological defects.  We discuss how the usual picture of axion
strings slicing up axion domain walls is modified by the inclusion of
the non-linearities of the true axion potential.  We find that rather
than the simple picture of axion strings destroying walls by punching
holes in them, unstable pseudo-breather solitons are formed which
decay to axitons.  In the final section we discuss some possible
physical consequences of very dense axion clumps.

\vspace{48pt}
\thesection{\centerline{\bf II. COSMOLOGICAL EVOLUTION OF THE AXION FIELD}}
\setcounter{section}{2}
\setcounter{equation}{0}
\vspace{12pt}
\def\lqcd{{{\Lambda_{\rm QCD}}}}
\def\Teq{{{T_{\rm EQ}}}}
\def\vphi{{{\vec{\phi}}}}

The axion story begins with PQ symmetry breaking.  This symmetry
breaking occurs when a complex scalar field $\vphi$ with non-zero PQ
charge develops a vacuum expectation value. This PQ symmetry breaking
can be modeled by considering a potential of the standard form
$V(\vphi)=\lambda\left(|\vphi|^2 - f_a^2/2\right)^2$.  The axion is
the resulting Nambu--Goldstone degree of freedom from spontaneous
breaking of the global symmetry.  After PQ symmetry breaking at $T\sim f_a$,
but before QCD effects are important, the axion is massless.  However
since the PQ symmetry is anomalous, it is broken explicitly by QCD
instanton effects, leading to a mass for the axion.  In general the
instanton effects respect a residual $Z_N$ symmetry, and the axion
develops a potential due to instanton effects of the form $ V(a) =
m_a^2(f_a/N)^2 [1-\cos(Na/f_a)]$.  The axion field is often
represented in terms of an angular variable $\theta\equiv Na/f_a$, and
if $\theta$ is taken as the dynamical variable, its potential for
$N=1$ is
\be
\label{pot}
V(\theta)=m_a^2(T)f_a^2(1-\cos\theta)\equiv\Lambda_a^4(T)(1-\cos\theta).
\ee

Because QCD instantons are large, with a size set by $\Lambda^{-1}_{\rm QCD}$,
their
effects are strongly suppressed at high temperatures.  So for $T\gg\lqcd$,
the axions are effectively massless.  For $T\gg\lqcd$, the temperature
dependence of the axion mass scales as
\be
\label{mascale}
m_a^2(T)=m_a^2(T_*) (T/T_*)^{-n};  \qquad    n=7.4\pm0.2 .
\ee

When the field $\theta(x)$ is created during the Peccei-Quinn symmetry
breaking phase transition at $T \sim f_a$, it should be uncorrelated
on scales larger than the Hubble radius at that time \re{ll90}.  As
the temperature decreases and the Hubble radius grows (in a
radiation-dominated Universe the Hubble radius grows as $R_H(T)\equiv
H^{-1}(T) \propto T^{-2}$), the field becomes smooth on scales up to
the Hubble radius.  This continues until $T =T_1 \sim 1$ GeV
when the axion mass ``switches on,'' i.e., when $m_a(T_1) \approx
3H(T_1)$, and the axion mass begins to become important in the
equations of motion.  Coherent axion oscillations then transform
fluctuations in the initial amplitude of the axion field into
fluctuations in the axion density.

Since the initial amplitude of the coherent axion oscillations on the
scale of the Hubble radius is uncorrelated, one expects that typical
positive density fluctuations on this scale will satisfy $\rho_a
\approx 2 \bar{\rho}_a$, where $\bar{\rho}_a$ is mean cosmological
density of axions \re{hr88}.  At the temperature of equal matter and
radiation energy density, $\Teq = 5.5 \,\Omega_a h^2 \, {\rm eV}$
\re{kt}, non-linear fluctuations will separate out as miniclusters with
$\rho_{\rm MC} \approx 10^{-14}\, {\rm g \, cm^{-3}}$ \re{hr88}.  The
minicluster mass will be of the order of the axion mass within the
Hubble radius at temperature $T_1$, $M_{\rm MC} \sim 10^{-9}\,
M_\odot$.  The radius of the cluster is $R_{\rm MC} \sim 10^{13}$cm,
and the gravitational binding energy will result in an escape velocity
of $v_e/c \sim 10^{-8}$. Note that the mean phase-space density of
axions in such a gravitational well is enormous: $\bar{n} \sim
\rho_a m_a^{-4}v_e^{-3} \sim 10^{48}f_{12}^4$, where $f_{12} \equiv f_a
/10^{12}\, {\rm GeV}$.

We will show below that due to non-linear effects, a substantial
number of regions at $\Lambda_{\rm QCD}>T>\Teq$ can have an axion
density orders of magnitude larger than $2\, \bar{\rho}_a$.  These
form because the non-linear effects in the axion potential lead to the
formation of pseudo-soliton objects we call axitons.

The axitons are not true solitons because the field coherently
oscillates inside the axiton.  The oscillations of the field lead to a
red-shift of the energy density of the field in an expanding Universe.
Quantitatively, axitons resemble breathers of the ($1+1$)-dimensional
sine-Gordon model.

Eventually the energy density of the axiton is red-shifted to
sufficiently small values of the axion field so that non-lineraities can
be neglected, and the axiton configuration is frozen in the
cosmological expansion as is any linear fluctuation. However the energy
contrast relative to the homogeneous background will be large.

Once an axiton forms, its energy density scales as $T^3$ for
$T > \Teq$, so we can write $\rho_{\rm axiton} \equiv 3\,
(1+\Phi) \Teq s/ 4$, where $\Phi$ depends upon the initial conditions
of the axion field, i.e., the misalignment angle {\em and its
gradients} at $T_1$. Here, $s$ is the entropy density, and $\Phi=0$
corresponds to the mean axion density. The energy density inside a
given fluctuation is equal to the radiation energy density at $T =
(1+\Phi) \Teq$.  At that time the self gravity of the fluctuation comes to
dominate, and if $\Phi\ga 1$ it separates out from the cosmological expansion,
collapses, and forms a minicluster with density\footnote{The factor of $140$
results from a detailed calculation.}
\begin{equation}
\label{rhofl}
\rho_{\rm MC} \simeq 140 \Phi^3 (1+\Phi) \bar{\rho}_a(\Teq) \approx 3 \times
10^{-14} \Phi^3(1+\Phi)\left(\Omega_ah^2\right)^4{\rm g~cm}^{-3},
\end{equation}
Even a relatively small increase in $\Phi$ is important because the final
density depends upon $\Phi^4$.

Ours is not the first proposal that non-linear effects can lead to
large values of $\Phi$.  One mechanism whereby
non-linear effects can lead to amplification of the axion density was
recognized in Ref.\ \re{mt86}.  In that analysis it was proposed that
some correlation regions can have values of $\Phi$ larger than one
because the closer the initial value of $\theta$ is to the top of the
axion potential, the later axion oscillations commence. However, this
effect alone is not very significant. If the closeness of the initial
angle to the top of the potential is parameterized by $\xi \equiv
(\pi-\theta_i)/\pi$, then for $\xi$ in the range $0.1 \la \xi \la
10^{-3}$, $\Phi-1 \approx 1.5 (\theta_i /\pi)^2 \xi^{-0.35}$, and
$\Phi$ is significantly larger than $1$ only for initial values very
finely tuned to the top of the potential
\re{kt93}. Moreover, the axion field is not exactly coherent on scales of the
Hubble radius, and even small fluctuations will spoil this simple picture.

Our scenario for the generation of axion miniclusters mainly depends
upon the interplay of the non-linear effects in the potential and
gradients in the axion field.  The interplay of these two effects will
lead to the formation of axiton configurations in the axion field.  At
temperatures $T \gg T_1$, the potential is negligible in the equations
of motion compared to the gradient terms which force the field to be
homogeneous on scales less than the Hubble radius.  At $T \ll
\Lambda_{\rm QCD}$, gradients can be neglected and one can treat the
evolution of fluctuations as that of a pressureless gas. Clearly,
around the QCD epoch when the potential just starts to become
important in the equations of motion the gradient terms are still
important, and since the initial amplitude can be close to $\pi$, the
non-linear nature of the potential is also important.  In order to
find the energy density profile at freeze out one has to trace the
non-linear inhomogeneous field evolution through the epoch $T\sim
T_1$.

\vspace{48pt}
\thesection{\centerline{\bf III. INHOMOGENEOUS AXION FIELD EVOLUTION}}
\setcounter{section}{3}
\setcounter{equation}{0}
\vspace{12pt}

\centerline{{\bf A. Equations of Motion}}

We start with deriving the equations of motion for the axion field in a form
suitable for numerical calculations.  In an expanding, spatially flat Universe
with scale factor $R(t)$, the equation of motion for the axion field takes the
familiar form
\be
\ddot{\theta} +
3\frac{\dot{R}}{R}\dot{\theta}-\frac{1}{R^2(t)}\bar{\Delta}^2\theta  +
m_a^2(t)\sin\theta=0,
\ee
where dot denotes time derivative and $\bar{\Delta}$ is the Laplacian with
respect to comoving coordinates $\bar{x}$.

Rather than cosmological time, it is convenient to work with a conformal-time
coordinate.  In a radiation-dominated Universe the conformal time is
proportional to the scale factor $R$.  Using $R$ as the independent variable,
the equation of motion is
\be
\frac{d^2\theta}{dR^2}+\frac{2}{R}\frac{d\theta}{dR}
-\frac{1}{\dot{R}^2}\frac{1}{R^2(t)}\bar{\Delta}^2\theta +
\frac{1}{\dot{R}^2}m_a^2(R) \sin\theta=0.
\ee
Using the Friedmann equation, along with the dependence of the
expansion rate upon $R$ in a radiation-dominated Universe, we can
express $\dot{R}^2$ in terms of the Hubble radius at some arbitrary
epoch (denoted by subscript $1$):
$\dot{R}^2=H^2R^2=H^2(R_1)R_1^4/R^2$. Now defining conformal time
$\eta$ as $\eta\equiv R/R_1$, the equation of motion is
\be
\theta'' + \frac{2}{\eta}\theta' - \frac{1}{H^2(R_1)R_1^2}
\bar{\Delta}^2\theta + \frac{\eta^2}{H^2(R_1)} m_a^2(R) \sin\theta=0,
\ee
where prime denotes $d/d\eta$.  We use Eq.\ (\ref{mascale}) to find that in
conformal time the mass evolves as $m_a^2(R)=m_a^2(R_1)\eta^n$.  We can use the
remaining freedom in the choice of $R_1$ to simplify the equation of motion by
making the choice $m_a^2(R_1)=H^2(R_1)$, i.e., $\eta=1$ corresponds to the
epoch when the inverse of the axion mass is equal to the Hubble radius.  The
equation of motion then takes the form
\begin{equation}
\theta''+{2\over \eta}\theta'-\Delta^2\theta + \eta^{n+2}\sin\theta =0,
\label{eq2}
\end{equation}
where  $\Delta$ is now the Laplacian with respect to comoving coordinates, $x
\equiv H(R_1)R_1\bar{x}$.    In other words, $x=1$ corresponds to the Hubble
radius at the epoch when the Hubble radius is the inverse of the axion mass.

The equation of motion can be written as a wave equation by the introduction of
the field $\psi \equiv \eta\theta$: \begin{equation}
\ddot{\psi}-\Delta^2\psi + \eta^{n+3}\sin (\psi /\eta )=0 \,\, .
\label{eq3}
\end{equation}
The equation of motion is finally in a form convenient for the study of the
evolution of the axion field during the epoch when the mass switches on.  In
Table I we give the scaling  with $\eta$ of several important physical length
and mass scales.  We next turn to the specification of the initial conditions.

\renewcommand{\arraystretch}{1.3}
\begin{table}[t]
\footnotesize{\hspace*{0.2in} Table I: The scaling of physical quantities with
conformal time $\eta$.  To find the scaling in coordinate distance, a
length must be divided by $\eta$, and a mass multiplied by $\eta$.}
\begin{center}
\begin{tabular}{l|l}
\hline \hline
$\phantom{XXX}$ TIME & $\phantom{XXX} t(\eta)=t(\eta=1)\eta^2$ 		\\
$\phantom{XXX}$ TEMPERATURE & $\phantom{XXX} T(\eta)=T(\eta=1)\eta^{-1}$\\
$\phantom{XXX}$ SCALE FACTOR & $\phantom{XXX} R(\eta)=R(\eta=1)\eta$ 	\\
$\phantom{XXX}$ AXION MASS   & $\phantom{XXX}
m_a(\eta)=m_a(\eta=1)\eta^{n/2}\quad n=7.4\pm 0.2 \phantom{XXX} $	\\
$\phantom{XXX}$ HUBBLE RADIUS $\phantom{XXXX}$ & $\phantom{XXX} R_H(\eta)
\equiv H^{-1}(\eta)=R_H(\eta=1)\eta^2$	\\
\hline \hline
\end{tabular}
\end{center}
\end{table}

\vspace{16pt}
\centerline{{\bf B. Initial Conditions}}

At $\eta \ll 1$ the potential term in Eq.\ (\ref{eq3}) can be neglected, and
the solution of the wave equation can be expressed simply as a sum of Fourier
harmonics. As usual, there will be two sums over frequency $\omega$: one sum
proportional to $\sin(\omega\eta)$ and one sum proportional to
$\cos(\omega\eta)$. In the decomposition of the $\theta$ field, terms like
$A(\omega) \sin (\omega \eta)/(\omega \eta)$ and  $B(\omega) \cos (\omega
\eta)/(\omega \eta)$ will appear. Assuming a finite amplitude for fluctuations
of $\theta$ (of order of several $\pi$) on scales larger than the Hubble radius
at the epoch of the Peccei--Quinn phase transition, we see that the
coefficients $B(\omega)$ must be proportional to $\Lambda_{\rm QCD}/f_a$, while
the coefficients $A(\omega)$ are of order unity.    In other words the terms
proportional to $\cos(\omega\eta)$ correspond to decaying modes on scales
larger than the Hubble radius and can be neglected. Finally, assuming that on
large scales the distribution for $\theta$  is white noise, we obtain
\begin{equation}
\theta = A \pi \sum_{ijk}{\sin (\omega \eta )\over \omega \eta } \sin (p_i x +
\varphi_{1ijk})\sin (p_j y + \varphi_{2ijk})\sin (p_k z + \varphi_{3ijk})  \,
\, ,
\label{inc}
\end{equation}
where $\varphi $'s are random phases and $\omega^2 =p_i^2 + p_j^2 +p_k^2$.
On scales larger than the Hubble radius the field distribution is frozen, while
modes smaller than the Hubble radius are redshifted away.

We numerically evolved this distribution starting from initial time $\eta =0.4$
in a box of size $L=4$ with periodic boundary conditions. There were $100^3$
grid points in the box. Each of the momenta in the field decomposition  took
six discrete values, $p_n = 2\pi n/L$, with $n=1, ..., 6$. So, in total there
were $3 \times 6^3$ random phases, each with values in the interval $0 <
\varphi < 2\pi$.

The final parameter to be chosen is the magnitude of $A$.  Recall that for
$N=1$, the axion potential is periodic with period $2\pi$. We will consider two
possibilities: $A=1$ and $A=2$.  For the case $A=1$ it is unlikely that domain
walls will form in a box of the size we study.  However for $A=2$ (the more
physical choice) domain walls are produced at about 1 per horizon. We will
present some results where $A=2$, but for the most part we will consider in
detail calculations with the $A=1$ initial condition, since we are interested
in the structure of density enhancements that are not associated with axion
domain walls.  So unless otherwise specified, our results will be for $A=1$.

The initial conditions are illustrated in Fig.\ 1 by a 2-dimensional slice
through the 3-dimensional box.  The height above the plane is proportional to
the axion energy density.   Since at this epoch the axions are relativistic, it
is convenient to scale their energy density by $\eta^4$.  The energy density
shown in Fig.\ 1 is scaled by $\eta^4/\bar{\rho}_a(\eta=3)$ where
$\bar{\rho}_a(\eta)$ represents mean axion energy density at a given $\eta$.
Note that the Hubble radius at this epoch ($\eta=0.4$) is $0.4$ in the units of
the figure, and the inverse of the axion mass is $75$ units.

\vspace{16pt}
\centerline{{\bf C. Results of Numerical Calculations}}
\vspace{12pt}

		\centerline{{\it 1. $(3+1)$-dimensional evolution}}

We first present the results of numerical calculations with $A=1$, where
density peaks arising from collapsing domain walls are filtered out so as to
isolate the effects due to axitons. In order to present the results of the
calculations we will take a two-dimensional slice through the three-dimensional
box, and plot the energy density as the height above the plane.  We have
analyzed the time evolution of the energy density in several different slices.
All of the slices generally look alike. The most important (and generic)
feature is the development of large-amplitude peaks. As the system evolves in
time, the peaks in the energy density, the axitons,  increase in magnitude and
become more compact.  We present the results in the $z=$ const plane, which
intersects the point with the  maximum energy density.  We emphasize that all
slices through the box are quantitatively similar. We normalize the
energy-density by comparing it to the energy density of a homogeneous axion
field, $\bar{\rho}_a(\eta)$, with initial amplitude equal to the {\em rms}
value of the misalignment angle, $\theta_{rms}=\pi /\sqrt{3}$.

In order to isolate the effect of the non-linearities in the axion potential,
we also evolve the same initial conditions with a harmonic axion potential,
$V(\theta) = m^2(T)f^2 \theta^2 /2$, and compare the evolution of the harmonic
potential model to the axion model.

The distribution of the axion energy density in the reference plane at time
corresponding to $\eta = 2$ is shown in Fig.\ 2a for the harmonic potential,
and in Fig.\ 2b for the axion potential. The maximum energy density peak that
picks the reference plane is clearly seen in Fig.\ 2b, its top portion is
choped off to fit overall the scale of the figure.

The distribution of the axion energy density in the reference plane at time
corresponding to $\eta = 3$ is shown in Fig.\ 3a for the harmonic potential,
and in Fig.\ 3b for the axion potential.   Again, the tops of the four peaks in
Fig.\ 3b are chopped off; their heights are in excess of $100$!  Of course the
peaks are only evident for the axion potential model.

Comparing Fig.\ 3b to Fig.\ 2b, we see that for the axion potential most of the
high magnitude peaks grow considerably in height and became thinner, while most
of the low amplitude peaks remain almost unchanged, i.e., they are in the
linear regime and consequently are frozen by the cosmological expansion. There
are some peaks (some even relatively high at $\eta=2$) which decreased in
amplitude. Those peaks represent the tales of the density clumps which reach
their maximum at some other value of $z$. All high density peaks contract in
the coordinate volume,  those which decreased in height simply moved out of
our reference plane. High density peaks do not develop in the evolution of the
harmonic potential, and the evolution proceeds as was assumed in the linear
analysis \ret{mt86}{hr88}.

There is insufficient resolution on this grid to proceed further in time with
the axion potential, but the harmonic potential can be evolved further.  In
Fig.\ 4 we present the result of the distribution of the axion energy density
in the reference plane at time corresponding to $\eta = 4$ for the harmonic
potential to demonstrate that as expected the evolution of the field in the
linear regime is frozen by $\eta=3$.  Note that the typical magnitude of the
peaks is about 2 for the harmonic potential.

There is a simple, heuristic explanation for the fact that non-linear effects
lead to the formation of high density peaks. The average pressure over a period
of homogeneous axion oscillations in the axion potential potential is
negative,\footnote{Of course the average pressure is dominated by relativistic
species at this time.  It is the pressure contributed by the axions that is
negative.} and is equal to $\pag\simeq - \Lambda_a^4(T) \theta_0^4 / 64$, where
$\theta_0$ is the amplitude of the oscillations \re{it86} (this formula is
valid for $\theta_0 \ll \pi$; as $\theta_0 \rightarrow \pi$, the field spends
more and more time near the top of the potential, and $\pag \rightarrow
-2\Lambda^4_a$). In other words, the axion self-interaction is attractive. The
larger the amplitude of oscillations inside the fluctuation, the more negative
the pressure inside, and consequently, fluctuations with excess axions will
contract in the comoving volume. In addition, matter with a smaller pressure
suffers less redshift in cosmological expansion.

Before continuing our exploration of the evolution of the peaks by means of a
1-dimensional calculation, we present some results of calculations with $A=2$,
where domain walls are much more likely to form than the above calculation with
$A=1$. The best way to illustrate the presence of domain walls is by a contour
graph, where the shading represents the amplitude of the axion energy density.
 We show a graph of the energy density distributions for the axion potential at
time $\eta =2$ with $A=2$ in Fig.\ 5a and compare it to a similar contour graph
for $A=1$ at $\eta=3$ in Fig.\ 5b.  In Fig.\ 5a two  shells of collapsing
domain walls are clearly visible in the lower left hand corner and in the upper
right hand corner.  Such configurations do not appear in Fig.\ 5b.  The density
peaks in Fig.\ 3b, the axitons, are not related to axion domain walls.

In order to learn the fate of the high density peaks, we have choosen one of
them in Fig.\ 3 and generated the corresponding spherically symmetric initial
conditions at $\eta=0.4$ and evolved it in time.  We now describe the result of
this calculation.

\vspace{12pt}
		\centerline{{\it 2. $(1+1)$-dimensional evolution}}

The axiton we choose to examine is the one near the center of the grid of Fig.\
3, with grid coordinates $\{2.24,1.92\}$ (the grid coordinate of the plane of
Fig.\ 3 in the $z$-direction is $1.76$---also near the center of the
3-dimensional box---and the axiton we chose is almost at its maximum in this
plane, having an absolute maximum at $z=1.80$).

The dependence of the axion field upon time at the reference point at the
center of this peak in our $(3+1)$-dimensional numerical calculation is shown
in Fig.\ 6 by the solid curve.   We can compare this evolution to the evolution
of a massless axion field with the same initial conditions since we are able to
calculate its evolution anaytically from the massless wave equation with
initial conditions given by Eq.\ (\ref{inc}).   The evolution of a massless
field at the reference point is shown in the Fig.\ 6 by the dashed line.

It is then straightforward to construct a spherically symmetric solution to the
massless counterpart of Eq.\ (\ref{eq3}) which has exactly the same time
dependence  as Fig.\ 6 in the center of the configuration.   We start with the
field decomposition, Eq.\ (\ref{inc}), substitute the values of the coordinates
of the given spatial point, and multiply each time harmonic by $\sin (\omega
r)/\omega r$.  So the dashed line in Fig.\ 6 also represents the time
dependence for a massless field at the reference point of the
($3+1$)-dimensional calculation and also at the center of a ($1+1$)-dimensional
spherically symmetric configuration.  (Away from the reference point even the
massless field will evolve differently in the ($1+1$)-dimensional calculation
and the $(3+1)$-dimensional calculation.)  We can use the resulting
configuration $\theta=\theta (\eta ,r)$ for generating spherically symmetric
initial conditions which will approximate the peak of our choice for the runs
with the actual axion potential.  The result of such a calculation with $A=1$
is shown in Fig.\ 6 by the dotted line.

This plot is instructive in the evaluation of the accuracy of the numerical
code.  At $\eta \la 1$ the axion field is approximately massless, and the
dotted curve is indistingushable from the dashed curve (there were $10^4$ grid
points in $r$-direction in the case of spherical symmetry) and the solid curve
deviates only very slightly (near the extrema) from the dashed curve.  This
suggests
that the use of $100$ grid points in each spatial direction in the
$(3+1)$-dimensional calculations is adequate.

{}From Fig.\ 6, we see that the axion mass effectively switches on at
a time $\eta \sim 1.3$. By this time the amplitude of the massless
field is greater than unity.\footnote{Note that around $\eta=2$ the
amplitude for the massless calculation is slightly larger than the
amplitude of the calculations using the true axion potential.  This is
because as the axion mass switches on the amplitude of the axion field
decreases.} This means that for the evolution of the field using the
axion potential the oscillations start in the non-linear regime
$\theta \ga 1$ in the region that will develop into a axiton.  We see
also that the non-linearity is strong enough to force the density peak
to collapse not only in coordinate space, but also in physical space
as well, since by the time $\eta \sim 3$ the amplitude of axion field
oscillations are in the non-linear regime and growing (this is somewhat
difficult to see in the figure). The rate of growth in the non-spherical case
is much slower (compare the solid and dotted lines).  This makes sense
because we expect a spherical collapse to lead to a denser central
region.

So in general, there are two competing effects in the evolution of the axion
field. 1) A contraction of the axiton due to the pressure difference,
leading to an increase in amplitude in the center, and 2) a decrease
in amplitude of the oscillations due to the expansion of the Universe.
We found that with a sufficiently large initial amplitude at the start
of oscillations, the first process wins, and the amplitude in the
center of the axiton increases to values $\theta \ga \pi$.  In the
opposite case, e.g., if the initial amplitude when oscillations
commence is in the linear regime, the amplitude monotonically
decreases in time.

In realistic axion models the axion mass does not continue to grow
with $\eta$, but saturates at its zero-temperature value around $T\sim
\lqcd$.  For axion masses in the allowed window this corresponds to
$\eta \ga 6$.  Due to the steep power-law dependence of the function
$m(\eta)$ in Eq.\ (\ref{mascale}), the period of the field
oscillations becomes very small by $\eta\sim 6$, and direct numerical
methods fail even in the case of spherical symmetry.  In order to
follow the evolution of the fluctuation up to freeze out we must
assume that the mass saturates at a smaller value of $\eta$.  This
would correspond to a larger value of $f_a$.  We have approximated the
procces of axion mass saturation by the the simple formula
\be
\label{saturate}
m^2(\eta)=m_a^2(\eta=1)\eta^n /[1+ (\eta /\eta_c)^{n}],
\ee
taking $\eta_c =3.5$.
This value of $\eta_c$ corresponds to too large value of $f_a \sim
4\times 10^{13}$GeV, which would give $\Omega_ah^2$ in excess of one.
However we expect that qualitatively the evolution of the axion field
will have the same basic properties for larger values of $\eta_c$
(smaller values of $f_a$).

We can vary the initial overall amplitude $A$ of our spherically
symmetric configurations.  This has the effect of spanning different
initial conditions of a well defined one-parameter family of axitons.
Moreover, varing $A$ is easier than choosing different peaks in Fig.\
3.

The time dependence of the field in the center of the fluctuation that
will develop into an axiton for $A=0.73$ is presented in Fig.\ 7, and
for $A=0.77$ in Fig.\ 8. In both cases the configuration collapses and
the amplitude of $\theta$ rapidly increases in the center, even
exceeding the value of $\pi$. This is followed by a period of several
rebounds.  An expanded view of the rebounds is shown in Fig.\ 9.
During each rebound (eight in total in both cases) relativistic axions
are emitted.  We can see the signature of axion emission by looking at
the radial profile of an axiton. In Fig.\ 10 we show the profile of
the axiton of Fig.\ 9 at three instants in time during one oscillation
period.  The emission of relativistic axions is seen in the outgoing
waves of Fig.\ 10.  The emission of relativistic axions reduces the
energy of the central configuration below some critical value, at
which point a pseudo-soliton, an axiton, is produced.

The radial coordinate $r$ is the spherical analogy to $x$. At the
values of $\eta$ in Fig.\ 10 the axion mass has saturated to
$m_a=3.5^{3.7}\eta \simeq 100\eta$.  Therefore in the units of Fig.\
10, the Compton wavelength of the axion is $0.01\eta^{-1}$, and the
axiton is obviously much larger than $m_a^{-1}$---it is indeed a
soliton-like configuration.

The axiton is a quasi-stable (on time scale $m_a^{-1}$) solution of the field
equations in an expanding Universe.  Since there are no absolutely stable
spherically symmetric breather-like solitons in flat space, in Minkowski
space-time an axiton configuration will gradually decay anyway without the
emission of axions present in the violent oscillations seen in Figs.\ 7 and 8.
In an expanding Universe the situation is different. Once the axiton enters the
linear regime it becomes frozen by the cosmological expansion, and behaves as a
clump of coherent field oscillations (or ultra-cold axions).

The final energy density profile of this configuration for the case $A=0.77$ is
shown in Fig.\ 11. At time $\eta =9$ (dotted line), outgoing secondary waves
are still seen in the tail of the configuration.  By time $\eta = 11$ there is
no evidence of outgoing radiation.  The amplitude of the energy density at
$r=0$ is $23.5$ at $\eta=9$ and $12.9$ at $\eta=11$ (the energy density in this
graph is not normalized to the homogeneous background).  It is clear that the
energy density in the center scales as $\eta^{-3}$ (e.g., $23.5/12.9 =
(11/9)^3$), confirming that the linear regime has been reached,  the
fluctuation is frozen, and the number of axions per comoving volume is
conserved.  The energy density of a homogenoeus background at $\eta = 10$ with
$\eta_c =3.5$ and initial amplitude equal to the {\it rms} value of $\theta$ is
0.85 in the units of the figure. Thus, the fluctuation of Fig.\ 11 has an
energy density contrast of 20.

Not all fluctuations that pass through the non-linear regime contract in
physical space.  For example, a sample spherical fluctuation for $A=0.70$ does
not collapse.  The corresponding energy density profiles of this fluctuation at
two moments of time are presented in Fig.\ 12 by the dashed lines.  This
should be compared to the solid lines, which are the energy density profiles
for a fluctuation with $A=0.73$ which does undergo collapse.  We see that the
slope of the energy density in the non-linear tail tends to a power law $\rho
\propto r^{-3}$ prior to the collapse.  This leads to an increase in field
amplitude in the center, while due to the overall expansion of the Universe,
the amplitude decreases.  For $A=0.73$, the first process wins for some period
of time, see Fig.\ 7, while for $A=0.7$ the general expansion dominates, and
the amplitude of the oscillations decreases monotonically.  However, the
decrease in amplitude is much slower than it would be with the harmonic
potential, and the final energy density contrast with $\eta_c=3.5$ and $A=0.7$
is $45$.

For comparison we also present in Fig.\ 12 the energy density profile of the
fluctuation with $A=0.77$ at $\eta =11$.  Remarkably, it has the same power-law
slope, $\rho \propto r^{-3}$, despite the fact that this profile represents a
fluctuation that has undergone ``violent oscillations'' accompanied by axion
emission (see Figs.\ 8, 10 and 11).

Since the axion interaction is attractive, one can expect that bound states of
axions can form.  One example of such a bound state is the well known
``breather'' solution in the ($1+1$)-dimensional sine-Gordon model. In ($3+1$)
dimensions this solution possesses planar symmetry and turns out to be unstable
with respect to fragmentation (we dicuss this further in the next section).  If
a spherically symmetric counterpart of the ``breather'' would exist in
Minkowski space-time, it would behave in an expanding Universe just as the
fluctuation shown in Fig.\ 11.  Thus the axiton is related to a spherically
symmetric breather.

Suppose we can extrapolate these results to the range of realistic axion
models, i.e., to larger values of $\eta_c$ corresponding to smaller values of
$f_a$.   Then we must consider the possibility of producing enormous density
contrasts. Indeed, both the increase in axion mass and the expansion of the
Universe adiabatically decreases the amplitude of axion oscillations in the
linear regime (or in the homogeneous state), so that at $T \la \Lambda_{\rm
QCD}$ the corresponding background energy density is about $ \bar{\rho}_a
\approx  \Teq T^3$, where $\Teq \sim 5.5\Omega_a h^2  {\rm eV}$ is the
temperature of equal radiation and axion energy density.  In the case of a
collapsing non-linear fluctuation, the final field configuration is the output
of non-linear dynamics. Let $\theta_L$ be the amplitude of field oscillations
in the axiton at the time when it enters the linear regime at $T_L=T_1/\eta_L$.
Then the corresponding energy density in the fluctuation will be at this time
about $\Lambda_{a}^4 \theta_L^2 $. The ratio of the axiton energy density to
the homogeneous background axion energy density will be
\be
1+\Phi \approx \Lambda_a^4\theta_L^2\eta_L^3/T_{\rm EQ}T_1^3.
\ee
Using the results from Figs.\ 7 and 8 ($\theta_L \sim 0.1$, and $\eta_L
>6),$\footnote{In any case, $\eta_L$ will be larger than $\eta_c$, the value of
$\eta$ where the axion mass saturates to its zero-temperature value [see Eq.\
(\ref{saturate})], and $\eta_L>6$ seems a very conservative estimate.} we
obtain $1 + \Phi \approx 10^4$ prior to gravitational decoupling of the
fluctuations from the cosmological expansion.

Although this possibility is exciting, a word of caution is necessary.
Non-linear dynamics is rather unpredictable, and one can not exclude the
possibility that at $\eta_c > 6$ all collapsing non-linear fluctuations
somehow dissipate, leaving very small $\theta_L$. Note also that
non-spherical configurations can evolve quite differently than the spherical
configurations.

The range of initial conditions which will lead to monotonic behavior of the
amplitude in the non-spherical case is expected to be wider. Our point of view
is that spectrum of energy density contrasts can span the entire range from
order $1$ up to of order $10^4$ or even larger.  However, at this time we have
nothing to say in regard to the number density of peaks as a function of its
amplitude.

So far we have neglected the presence of other non-linear structures which can
be formed by the axion field during the QCD epoch, namely axion domain walls
and walls bounded by strings. We now turn to the question of their fate and
their contribution to the dark matter distribution.

\vspace{48pt}
\thesection{\centerline{\bf IV. AXION BREATHERS}}
\setcounter{section}{4}
\setcounter{equation}{0}
\vspace{12pt}

 In general, there are four sources of cosmic axions.  The first
source is thermal axions \re{THERMAL}.  The second source, related to
the initial misalignment of the axion degree of freedom from its true
minimum, was discussed in the previous sections.  We will refer to
this source of axion energy density as the misalignment energy
density.  The third source is the decay of cosmic axion strings
\ret{rd86}{hs87}.  In Ref.\ \re{rd86}, it was found that the energy density
resulting from this process is two orders of magnitude larger than the
misalignment energy density, while in the estimate of Ref.\ \re{hs87},
the energy density from the decay of cosmic strings is comparable to
the misalignment energy density.  At $T \sim T_1$ the decay of strings
will also produce an inhomogeneous axion field.  While we can not
describe the initial configuration emerging from string decay by the
distribution of Eq.\ (\ref{inc}), we expect that the attractive
non-linear self interaction will also play a role here, and the
evolution will proceed along the lines described in Sec.\ III and will
result in high density peaks.  The fourth potential source of axions
is related to the collisions and subsequent disappearance of axion
domain walls.  In this section we discuss this last process.

In most cases a network of vacuum domain walls is a cosmological
disaster, since they soon come to dominate the energy density of the
Universe \re{vko75}.  Fortunately, in the $N=1$ axion model domain
walls are effectively unstable and this problem is avoided. The
process by which the domain walls disappear is through collisions of
the string network with the walls.  The usual assumption is that when
a string loop (with a wall on the inside) hits a large wall the two
pieces of wall annihilate and a vacuum hole is produced in the large
wall.  Since the surface energy of the hole is smaller than the
surface energy of the wall, the vacuum hole expands and devours the
wall.  For an infinite domain wall there will be roughly one hole in
the wall per Hubble radius, so in a couple of Hubble times the holes
quickly overlap and the wall disappears.  Domain walls of finite size
(size smaller than the Hubble radius) form closed surfaces and shrink
by themselves.  An oversimplified point of view would be that all of
the energy released in the disappearance of the domain walls is
transferred to relativistic axions, which subsequently redshift away
and would become an insignificant source of axion energy density.

We shall argue here that the hole in a domain wall formed by a
string-loop intersection is not vacuum, but rather consists of a bound
state of two pieces of domain wall (which in the simplified scenario
annihilated each other) corresponding to a generalization of
the ``breather'' solution of the ($1+1$)-dimensional sine-Gordon model
\re{rr82}.  That is, the vacuum wall network is transformed into a
breather wall network. The breather wall effectively evolves as a dust
wall rather than a domain wall, so it represents cold dark matter.

We consider here the axion field in Minkowski space-time with planar symmetry
as a function of two coordinates, time $x_0$ and one spatial direction, $x_1$.
It is convenient to introduce the dimensionless variables $t \equiv x_0 \, m_a$
and $z \equiv x_1 m_a$. The relevant breather solution to the equation of
motion $\ddot{\theta} -\theta^{\prime \prime} + \sin \theta =0$ (dot denotes
$d/dt$ and prime denotes $d/dz$) has the form \re{rr82}
\begin{equation}
\label{breathers}
\theta_\upsilon (t, z) = 4\, \arctan \left[ {\sin (t \,/\, \tau) \over \upsilon
\cosh (z\, /\, L) }\right]   \, ,
\label{br}
\end{equation}
where $\tau \equiv \sqrt {1 + \upsilon^2}/\upsilon $ and $ L \equiv
\sqrt {1 + \upsilon^2}$. One can interpret this solution as a bound
state of two static domain walls (or kinks), $\theta_{\rm kink} = \pm
4\, \arctan [\exp (z)]$. The free parameter $\upsilon$ of the breather
is related to the binding energy.  Larger $\upsilon$ corresponds to
larger binding energy of the kinks.  In terms of the spatial energy
density distribution, the breather looks like a domain wall with an
effective width $L$, but unlike the usual static domain wall, the
field coherently oscillates with period $\tau$ in the breather.  When
$\upsilon \rightarrow 0$ the period tends to infinity, and when
$\upsilon \rightarrow \infty$ the field oscillates with a frequency
equal to the axion mass.  The width of the breather scales in the
opposite way with $\upsilon$: as $\upsilon
\rightarrow 0$ the width is $m_a^{-1}$, and the width grows proportional to
$\upsilon $ at large $\upsilon$.

When considering wall like structures, it is convenient to introduce a surface
stress-energy tensor of the wall, $S^\mu_{~\nu} \equiv \int_{-\infty}^\infty \,
T^\mu_{~\nu} \, dz$. While all the components of $T^\mu_{~\nu}$ in the breather
solution oscillate with time, the surface energy density is constant:
\begin{equation}
S^0_{~0} = 16f_a^2 m_a/\sqrt{1+\upsilon^2} \, ,
\label{s00}
\end{equation}
and  $S^\mu_{~\nu} $ is a diagonal tensor.  The spatial components of
$S^\mu_{~\nu}$ are oscillating functions. However, when considering the
macroscopic properties of a wall, the relevant quantities are averages over an
oscillation period. Upon averaging over an oscillation period $\langle
S^z_{~z}\rangle=0 $, as it must be for a wall of any nature \re{bkt87}.  For
the time-averaged surface tension we find
\begin{equation}
\langle S^i_{~j}\rangle =8f_a^2 m_a \left[ 2 \left( \sqrt{1+\upsilon^2}
-\upsilon \right) - {\upsilon \over 1+\upsilon^2} \right]\delta^i_{~j} \equiv
S \delta^i_{~j} \, .
\label{sij}
\end{equation}
As $\upsilon \rightarrow 0$, the stress-energy tensor tends to the
vacuum stress tensor, with $S^0_{~0}=S$, where $S^0_{~0}$ is twice the
energy density of a single kink.  However, at large $\upsilon$, we
have $S^0_{~0} \approx 16\,\Lambda_a^4 \, /m_a\, \upsilon$ and $S
\approx 8\,\Lambda_a^4 \, /m_a\,\upsilon^3$.  With increasing
$\upsilon$, this tends to the stress-energy tensor of a dust shell.
So in the expansion of the Universe, the surface density of the
breather wall has to decrease and $\upsilon$ has to grow. Using Eqs.\
(\ref{s00}) and (\ref{sij}), we obtain as a solution to the planar
wall equations of motion \re{bkt87}
\begin{equation}
S^0_{~0}(R) = {32 f_a^2 m_a \, R^2 \over R^4 + 1} \, ,
\label{soa}
\end{equation}
where we assumed constant $m_a$ and have have normalized the scale
factor in such a way that $R=1$ at the moment when $\upsilon=0$. Note
that the number of axions per unit area is conserved at large $R$.
Despite the fact that the breather is a bound state, its surface
energy density decreases in expansion, exactly as the energy density
of the solution presented in Fig.\ 11.

We can visualize the formation of the breather network in the
following scenario.  When domain walls form at $T \sim T_1$, every
string loop develops a wall inside (an ``antiloop'' develops a wall
outside).  When a string hits a large segment of wall, the
intersection region will not be empty, but will be a bound state of
two domain walls.  In the idealized approximation of planar symmetry,
the bound state will correspond to the breather solution of Eq.\
(\ref{br}). However, since the perfectly planar situation is
unrealistic, the question arises whether breather walls are stable.

To answer this question we numerically integrated the axion equations
of motion in Minkowsi space-time with initial conditions corresponding
to a perturbed breather wall. We evolved an axisymmetric configuration
$\theta = \theta (t,z,r)$, which initially corresponded to the breather
field distribution of Eq.\ (\ref{breathers}) with $\upsilon =
\upsilon (r)$.  The value of $\upsilon$, and the corresponding pressure, was
larger in the center (note that Eq.\ (\ref{sij}) corresponds to a
system with negative pressure).  In a sense, this configuration
coresponds to a bubble of new phase of lower energy density, and it is
expected to expand.  The question is whether the field inside the
``bubble'' will tend to a breather solution with a new constant value
of $\upsilon$ as the boundary of the bubble propagates outward. We
have found that this does not occur: the breather wall {\it is
unstable}.  However, the energy density in the breather does not
dissipate, but the breather fragments into clumps very similar to
those discussed in Sec.\ III.C.  This result is not unexpected in view
of the attractive nature of the axion self-interaction.  The energy
density profile in the $r$-direction is presented in Fig.\ 13 at
several moments of time.

Our conclusion in this section is that the decay of the axion domain wall
network can provide yet another channel for axiton production in the axion
distribution.

\vspace{48pt}
\thesection{\centerline{\bf V. DISCUSSION}}
\setcounter{section}{5}
\setcounter{equation}{0}
\vspace{12pt}

In principle, all axion miniclusters could be relevant to
laboratory axion search experiments, since even for $\Phi$ as
small as 1, the density is $10^{10}$ times larger than the local
galactic halo density [see Eq.\ (\ref{rhofl})].  Moreover, as we have noted
already, the energy density in an axiton after it separates out from the
general expansion will be $\Phi^4$ times larger than the energy
density at $\Teq$.  For example, a rather moderate density
contrast of $\Phi=30$ at $\Lambda_{\rm QCD}>T>\Teq$ will correspond to
roughly an additional factor of $10^6$ in the energy density of
the axiton at $T\ll \Teq$.

The probability of a direct encounter with a minicluster is small.  Let's
assume that all of the axions end up in miniclusters of mass $10^{-9}M_\odot$,
density $10^{-14}$g cm$^{-3}$, and radius $4\times 10^{12}$cm.  Using a local
halo mass density of $5\times 10^{-25}$g cm$^{-3}$ would give a minicluster
number density of 7,000,000 pc$^{-3}$.  With a typical velocity of 250 km
s$^{-1}$ the encounter rate would be 1 per 25 million years.  Although the
signal in an axion detector from a close encounter with a minicluster would be
enormous, it might be a long wait.  So the interesting question arises, could
there be any other astrophysical consequences of very dense axion clumps?
Below we shall discuss the
possibility of ``Bose star'' formation in axion miniclusters.

The physical radius of an axiton at $\Teq$ is larger by many
orders of magnitude than the de Broglie wavelength of an axion
in the corresponding gravitational well.  Consequently, the
gravitational collapse of the axion clump and subsequent
virialization can be described in the usual terms of cold dark
matter particles.  In a few crossing times some equilibrium
distribution (presumably close to an isothermal distribution) of
axions in phase space will be established. It is remarkable that
in spite of the apparent smallness of axion quartic
self-couplings, $|\lambda_a| = (f_\pi/f_a)^4 \sim 10^{-53}
f_{12}^{-4}$, the subsequent relaxation in an axion minicluster
due to $2a \rightarrow 2a$ scattering can be significant as a
consequence of the huge mean phase-space density of axions
\re{it91}. In the case of Bose-Einstein statistics the inverse
relaxation time is $(1+\bar{n})$ times the classical expression,
or $\tau_R^{-1} \sim \bar{n}\, v_e \sigma \rho_a /m_a $, where
$\sigma$ is the corresponding cross section. For particles bound
in a gravitational well, it is convenient to rewrite this
expression in the form \re{it91}
\begin{equation}
\tau_R  \sim  m_a^7 \lambda_a^{-2}\rho_a ^{-2}v_e^2 .
\label{rl}
\end{equation}
The shallower the gravitational well for a given density of
axions, the larger the mean phase space density, and
consequently the smaller the relaxation time due to the $v_e^2$
dependence in Eq.\ (\ref{rl}). Note also the dependence of the
inverse relaxation time upon the square of the particle density.

The relaxation time (\ref{rl}) is smaller then the present age
of the Universe if the energy density in the minicluster
satisfies
\begin{equation}
\rho_{10}  >  10^{6} v_{-8}\sqrt{f_{12}},
\label{rt}
\end{equation}
where $\rho_{10} \equiv \rho /(10 \, {\rm eV})^4$ and $v_{-8}
\equiv v_e /10^{-8}$. If this occurs, then an even denser core
in the center of the axion cloud should start to form. An
analogous process is the so-called gravithermal instability
caused by gravitational scattering.  This was studied in detail
for star clusters, where the ``particles" obey classical
Maxwell--Boltzmann statistics.  Axions will obey Bose--Einstein
statistics, with equilibrium phase-space density $n(p) = n_{\rm
cond} + [ e^{\beta E} -1 ]^{-1}$, containing a sum of two
contributions, a Bose condensate and a thermal distribution. The
maximum energy density that non-condensed axions can saturate is
$\rho_{\rm ther} \sim m_a^4 v_e^3 $, which corresponds to
$\bar{n}_{\rm ther} \sim 1$. Consequently, given the initial
condition $\bar{n} \gg 1$, one expects that eventually the
number of particles in the condensate will be comparable to the
total number of particles in the region if relaxation is
efficient.  Under the influence of self gravity, a Bose star
[\ref{it86}, \ref{rb69}, \ref{ss91}] then forms \re{it91}. One can
consider a Bose star as coherent axion field in a gravitational
well, generally with non-zero angular momentum \re{it86}.

Comparing Eqs.\ (\ref{rhofl}) and (\ref{rt}), we conclude that
the relaxation time is smaller than the present age of the
Universe and conditions for Bose star formation can be reached
in miniclusters with a density contrast $\Phi \ga 30$ at the QCD
epoch.

Under appropriate conditions stimulated decays of axions to two
photons in a dense axion Bose star are possible
[\ref{it86}, \ref{it87}] (see also \re{kw90}), which can lead to the
formation of unique radio sources---axionic masers. In view of the
results of this paper we conclude that the questions of axion
Bose star formation, structure, and possible astrophysical
signatures deserve detailed study.

In conclusion, we have presented a 3-dimensional numerical study
of the evolution of inhomogeneities in the axion field around the
QCD epoch, including for the first time important non-linear
effects.  We found that the non-linear effects of the attractive
self-interaction can lead to a much larger density of axions in
miniclusters than previously estimated. Large amplitude density
contrasts form solitons we call axitons, and resemble the
bound-state ``breather'' solutions of the ($1+1$)-dimensional
sine-Gordon model.  The increase in the axion density may be
sufficiently large that axion miniclusters formed by the fluctuations might
exceed the critical density necessary for them to relax to form Bose stars.

\vspace{18pt}

\centerline{\bf ACKNOWLEDGMENTS}

EWK and IIT are supported by the DOE and NASA under Grant NAGW--2381.

\begin{picture}(400,50)(0,0)
\put (50,0){\line(350,0){300}}
\end{picture}

\vspace{0.25in}

\def\labelenumi{[\theenumi]}
\frenchspacing
\def\prl{{{\em Phys. Rev. Lett.\ }}}
\def\prd{{{\em Phys. Rev. D\ }}}
\def\pl{{{\em Phys. Lett.\ }}}
\begin{enumerate}
\item \label{pq}
R. D. Peccei and H. Quinn, \prl {\bf 38}, 1440 (1977) and
 Phys. Rev. {\bf 16}, 1791 (1977); S. Weinberg, \prl
{\bf 40}, 223 (1978); F. Wilczek, \prl {\bf 40}, 279  (1978).
\item \label{others}
Other solutions to strong CP problem include vanishing of one of a
quark mass, see e.g., D. B. Kaplan and A. Manohar, \prl {\bf 56}, 2004
(1986); models where an otherwise exact CP is either softly or
spontaneously broken, see A. Nelson, \pl {\bf B136}, 387 (1984) and S.
M. Barr, \prl {\bf 53}, 329 (1984); or through the action of
wormholes, see B. Nielsen and M. Ninomiya, \prl {\bf 62}, 1429 (1989),
K. Choi and R. Holman, \prl {\bf 62}, 2575 (1989), J. Preskill, S. P.
Trivedi, and M. Wise, \pl {\bf 223}, 26 (1989).
\item \label{global}
R. Holman, S. H. Hsu, T. W. Kephart, E. W. Kolb, R.
Watkins, and L. M. Widrow, \pl {\bf B282}, 132 (1992); S. Barr  and D. Seckel,
\prd {\bf 46}, 539 (1992);  M. Kamionkowski and J. March-Russell, \pl {\bf
B282}, 137 (1992).
\item \label{ac1}
D. Dicus, E. Kolb, V. Teplitz and R. Wagoner, \prd {\bf 18},
1829 (1978) and \prd {\bf 22}, 839  (1980); M. Fukugita, S.
Watamura and M. Yoshimura, \prl {\bf 48}, 1522  (1982)
and \prd {\bf 26}, 1840 (1982).
\item \label{ac2}
G. Raffelt and D. Seckel, \prl {\bf 60}, 1793  (1988);
M. S. Turner, \prl {\bf 60}, 1797 (1988); R. Mayle et al., \pl {\bf 203B}, 188
(1988); T. Hatsuda and M. Yoshimura, \pl {\bf 203B}, 469  (1988).
\item \label{cc}
J. Preskill, M. Wise and F. Wilczek, \pl {\bf 120B}, 127 (1983);
L. F. Abbott and P. Sikivie, \pl {\bf 120B} , 133 (1983); M. Dine
and W. Fischler, \pl {\bf 120B}, 137 (1983).
\item \label{mt90}
For recent reviews, see M. S. Turner, {\em Phys. Rep.} {\bf C197}, 67 (1990);
G. G. Raffelt, {\em Phys. Rep.} {\bf C198}, 1 (1990).
\item \label{mt86}
M. S. Turner, \prd {\bf 33}, 889 (1986).
\item \label{hr88}
C. J. Hogan and M. J. Rees, \pl {\bf B205}, 228 (1988).
\item \label{ps83}
P. Sikivie, \prl {\bf 51}, 1415 (1983).
\item \label{kt93}
E. Kolb and I. I. Tkachev, \prl {\bf 71}, 3051 (1993).
\item \label{it91}
I. I. Tkachev, \pl {\bf B261}, 289 (1991).
\item \label{ll90}
This does not necessarily require the reheating temperature after inflation to
be higher than $f_a$, since inflation itself can produce strong fluctuations in
the axion field as discussed in
A. D. Linde and D. H. Lyth, \pl {\bf B246}, 353 (1990);
D. H. Lyth and E. D. Stewart, \prd {\bf 46}, 532 (1992).
\item \label{kt}
E. W. Kolb and M. S. Turner, {\em The Early Universe}, (Addison-Wesley, Redwood
City, Ca., 1990).
\item \label{gpy81}
D. Gross, R. Pisarski, and L. Yaffe, {\em Rev. Mod. Phys.} {\bf 53}, 43 (1981).
\item \label{it86}
I. I. Tkachev, {\em Sov. Astron. Lett.} {\bf 12}, 305 (1986).
\item \label{many}
Another consideration is that a typical minicluster mass is $10^{-9}M_\odot$.
If axions are the dark matter in our galaxy, then there are of order $10^{20}$
miniclusters in our galaxy.  Even if we have been too optimistic by several
orders of magnitude about the probability of initial configurations that lead
to large density contrasts, there will still be a sizeable number of them in
our galaxy.
\item \label{THERMAL}
For a discussion of thermal axion production, see \re{kt}.
\item \label{rd86}
R. Davis, \pl {\bf B180}, 225 (1986).
\item \label{hs87}
D. Harari and P. Sikivie, \pl {\bf 195}, 361 (1987).
\item \label{vko75}
 M. B. Voloshin, I. Yu. Kobzarev, and L. B. Okun', {\it Sov. J. Nucl. Phys.}
{\bf 20}, 644 (1975);
\item \label{rr82}
R. Rajaraman, {\it Solitons and Instantons}, (Elsevier Science Publishers, New
York, 1982).
\item \label{bkt87}
V. A. Berezin, V. A. Kuzmin, I. I. Tkachev, \prd {\bf 36}, 2919 (1987).
\item \label{rb69}
R. Ruffini and S. Bonozzola, {\em Phys. Rev.} {\bf 187}, 1767 (1969);
J. D. Breit, S. Gupta, and A. Zaks, \pl {\bf B140}, 329 (1984);
M. Gleiser, {\em Phys. Rev. D} {\bf 38}, 2376 (1988);
P. Jetzer, {\em Nucl. Phys.} {\bf B316}, 411 (1989).
\item \label{ss91}
E. Seidel and W.-M. Suen, \prl {\bf 66}, 1659 (1991).
\item \label{it87}
I. I. Tkachev, \pl {\bf B191}, 41 (1987).
\item \label{kw90}
T. W. Kephart and T. J. Weiler, preprint VAND-TH-90-2, 1990 (unpublished).
\end{enumerate}

\end{document}